# The "intrinsic helicity" of the elementary particles and the proof of the spin-statistics theorem.

## Enrico Santamato[1*] and Francesco De Martini[2]

[1] Dipartimento di Fisica, Università Federico II, Napoli, 80126 Italy.

[2] Accademia dei Lincei, Roma, 00165 Italy

*Correspondence to  enrico.santamato@na.infn.it

francesco.demartini@uniroma1.it

**Abstract.** The traditional Standard Quantum Mechanics is unable to solve the Spin-Statistics problem, i.e. to justify the utterly important "Pauli Exclusion Principle". We show that this is due to the non completeness of the standard theory due to an arguable conception of the spin as a vector characterizing the rotational properties of the elementary particles. The present Article presents a complete and straightforward solution of the Spin-Statistics problem on the basis of the "Conformal Quantum Geometrodynamics", a theory that has been proved to reproduce successfully all relevant processes of the Standard Quantum Mechanics based on Dirac's or Schrödinger's equations, including Heisenberg's uncertainty relations and nonlocal EPR correlations. When applied to a system made of many identical particles, an additional property of all elementary particles enters naturally into play: the "intrinsic helicity". This property determines the correct Spin-Statistics connection observed in Nature.

______________________________________________________________

## 1. Introduction

"*Everyone knows the Spin-Statistics theorem but no one understands it*" [1]. This puzzle represents the dramatic failure of the otherwise always successful Standard Quantum Mechanics (SQM) in a context of tremendous, cosmological relevance since it involves the very exitence of atoms, of ourselves, of the entire Universe. In the words of Richard Feynman: *"[It has been shown] that spin and statistics must necessarily go together but we have not been able to find a way of reproducing his arguments on an elementary level. It appears to be one of the few places in Physics where there is a rule which can be stated very simply but for which no one has found a simple and easy explanation. [….]This probably means that we do not have a complete understanding of the fundamental principles involved"* [2]. In the last decades a vast literature has grown about the Spin-Statistics Connection (SSC) [1–5]. In particular, attempts to model the quantum spin as a rotating frame attached to a point particle have appeared in the literature . In particular it is worth quoting here the work by Arthur Jabs who introduced the model of the ratchet gyroscope with one allowed sense of rotation around the proper axis of the particle [6,7].

In the present Article we shall present the explanation required by Feynman by introducing a novel, fundamental property of all elementary particles, the "intrinsic helicity" as a necessary theoretical completion to the Standard Quantum Mechanics, i.e. the one developed over nearly a century on the tracks of Bohr, De Broglie, Schrödinger, Heisenberg and Dirac. The essential clue of the present demonstration resides on an insightful and complete description of the "spin". Indeed this quantum object has always been considered by SQM and by Quantum Field Theory to consist of *axial vector* [3] on the basis of a general understanding of the processes underlying all known experiments in Physics involving spin interactions. One should consider that in reality all these experiments involve only mutually interacting "*magnetic moments*", µ − taken as axial vectors − or magnetic moments interacting with magnetic fields, external or internal to the structure (e.g. atoms) to which the spin belongs. In the present article we suggest that the only experiment in Physics where the simple vector picture appears to fail − and were the presumption of the magnetic interaction may play a misleading role – may be precisely the "Spin-Statistics connection". Indeed, the correct concept to be adopted for the spin lies down on its obvious and fundamental essence as the *angular momentum of a rotating frame* subjected to kinematic constraint, which renders it similar to a ratchet gyroscope, where the rotation around its proper axis $\zeta$ can have one sense (e.g. counter-clockwise) for any specific quantum particle with rotation angle $\gamma$ always increasing ($d\gamma/dt > 0$). This representation of spin imposes an additional fundamental constant property to be attached to each elementary particle in Physics, as mass and charge: the "<u>intrinsic helicity</u>" $s_\zeta$, i.e. the component of the angular momentum of the particle frame along its proper axis $\zeta$. As shown by all standard Texts of quantum mechanics, as for instance in [8], the concept of intrinsic helicity is *extraneous* to the lexicon of the SQM where one considers only the three components $s_x$, $s_y$, $s_z$ along the axes $x,y,z$ of the laboratory frame. In the following we will call "quantum spin" the object described by the components $s_x$, $s_y$, $s_z$ (or their quantum operators), considered by SQM and proportional to the corresponding magnetic moments. The quantity $s_\zeta$ will be considered instead as a novel, independent physical quantity bearing several relevant dynamical properties. For instance, although $s_\zeta$ has the same origin as $s_x$, $s_y$, $s_z$, it is a *conserved,* unmodifiable quantity whose value is not changed by any known force or potential acting on the particle. On the contrary, it is well known that $s_x$, $s_y$, $s_z$ can be modified by external fields so to take any value of their spectrum from $-\hbar s$ to $+\hbar s$, with $s$ integer or half integer. Nevertheless, although quantum states with simultaneously fixed values of $s_x$, $s_y$, $s_z$ do not exist, owing to the non commutativity of the corresponding operators, $s_\zeta$ retains its constant value together with each one of the spin components $s_x$, $s_y$, $s_z$ as shown in [8]. For simplicity and without loss of generality, we may orient the proper $\zeta$ axis of the particle frame so to have $s_\zeta = +\hbar s$ and then refer to the constant "*spin value*" $s$ as an intrinsic property of the particle, i.e. the "intrinsic helicity".

In the present work, we claim that the observed quantum statistics for fermions and bosons and Pauli's exclusion principle are determined precisely and exclusively by the properties of the "intrinsic helicity", $s_\zeta$ of the identical particles. The fact that $s_\zeta$ is extraneous to SQM structure justifies the failure of that theory in the context of the SSC, as stated by the first sentence reported at the beginning of the present Article. In other words, we contend that SQM cannot be considered a complete theory within the conceptual context of the Pauli Principle. On the other hand, we find impossible to fit coherently the new particle's property within the well established structure of SQM. A straightforward solution to the SSC problem is attained in the context of the "Conformal Quantum Geometrodynamics" (CQG), a modern approach to Quantum Mechanics (*9–16*). The intrinsic helicity is found to a fit in the CQG theory in a coherent and natural fashion. For instance, the constraint of one rotation sense around $\zeta$ is not assumed *ad hoc*, as done in [6,7], but emerges as a consequence of first principles.

## 2. The Conformal Quantum Geometrodynamics

In what follows we outline some of the most fundamental statements and definitions of the CQG for the sake of clarity and completeness. The CQG describes a quantum system by means of two real valued scalar functions $S(q,t)$ and $\rho(q,t)$ of the generalized coordinates $q^i$ ($i = 1,…,n$) and time, which fully determine the *quantum state* of the system. In the nonrelativistic limit, the fields $S(q,t)$ and $\rho(q,t)$ obey the following set of coupled partial differentuial equations

$$-\partial_t S = \frac{1}{2m} g^{ij}(\partial_i S - a_i)(\partial_j S - a_j) + V + \frac{\xi\hbar^2}{m} R_W \qquad (1)$$

$$\partial_t \rho + \frac{1}{\sqrt{g}} \partial_i \left( \sqrt{g} \rho g^{ij}(\partial_j S - a_j) \right) = 0 \qquad (2)$$

where the $a_i(q,t)$ and $V(q,t)$ represent vector and scalar external fields applied to the system, respectively, and sum over repeated indices is intended. The CQG assumes that the configuration space of the system is a metric space with metric tensor $g_{ij}(q)$ and with affine connections so that the change of the length $l$ of a vector in an infinitesimal parallel transport from $q^i$ to $q^i + dq^i$ is given by Weyl's rule $\delta l = -2l\phi_i dq^i$ [18], with Weyl's vector $\phi_i = -\left(\frac{1}{n-2}\right)\partial_i \ln \rho$. The Weyl connection of CQG is then integrable with potential $\rho$. The introduction of Weyl's connections renders the theory covariant with respect conformal changes of the metric tensor ($g_{ij} \to \lambda g_{ij}$) and gauge change $\phi_i \to \phi_i - \frac{1}{2}\partial_i \ln\lambda$ for arbitrary function $\lambda(q)$. This additional symmetry, however, is of relevance only within the relativistic approach, and can be neglected in the present non relativistic context. An important consequence of the Weyl parallel transport law for vectors is that the configuration space turns out to be curved even if the metric $g_{ij}$ is flat as in the Minkowski or Euclidean spaces. In facts, the Weyl scalar curvature $R_W$ is given by

$$R_W(q) = R_g + \left(\frac{n-1}{n-2}\right)\left[\frac{g^{ij}\partial_i\rho\partial_j\rho}{\rho^2} - \frac{2g^{ij}\partial_i\left(g^{ij}\partial_j\rho\right)}{\rho\sqrt{g}}\right] \qquad (3)$$

where $R_g$ is the Riemann curvature calculated from the Christoffel symbols of $g_{ij}$ and $n > 2$ is the dimensionality of the system configuration space. Therefore Eqs. (1) and (2) are coupled through $R_W$ and must be solved consistently. The fundamental result of CQG is that the Weyl's curvature is the origin of all quantum effects [9,11,14], including Heisenberg's uncertainty [17] and EPR correlations of spin ½ pairs [13,19]. Here we take for granted that the CQG fully reproduces the SQM and focus on the problem of the Spin-Statistic Connection. The SSC is a global property of the system while Eqs. (1) and (2) are local. We then discuss preliminarily some global properties of the fields $\rho(q,t)$ and $S(q,t)$. The CQG assumes that these two fields are single valued functions of the system configurations[1]. The uniqueness of $\rho$ is needed to have vectors with well defined length. In facts, if we parallel transport a vector along a loop in the considered region **X** of the configuration space, its length recovers its initial value, because $\oint \delta l/l = 0$, the Weyl's connection being integrable with uniquely defined potential $\rho$. The uniqueness of $S$ implies, for any loop in the region

---

[1] This property cannot be extended to the whole configuration space, in general, so we restrict our considerations to some region **X** where $\rho(q,t)$ and $S(q,t)$ are uniquely defined.

**X** under study (at any fixed time $t$)[2]

$$\oint dS = \oint p_i dq^i = 0 \tag{4}$$

where we set $p_i = \partial_i S$. On the other hand, if we call $H(q, \partial_i S)$ the right hand side of Eq.(1), we may identify this equation as the Hamilton-Jacobi Equation (HJE) of a mechanical system with Hamiltonian function $H$ and corresponding Lagrangian $L$ that in our case is given by

$$L(q,\dot{q},t) = \frac{1}{2} m g_{ij} \dot{q}^i \dot{q}^j + a_i \dot{q}^i - \frac{\xi \hbar^2}{m} R_W(q) - V(q) \tag{5}$$

The field $S(q,t)$ can be considered as the action function of the system. As it is well-known, condition (4) is necessary for the line integral $\int L dt$ to afford a strong minimum for fixed end-points in **X** [20]. In particular, according to Carathéodory, this condition allows to construct the so-called "complete figure" (*vollständige Figur*) of the calculus of variations. This figure is formed by the family of geodesically equidistant surfaces $S(q,t) = $ const. which are intersected by a canonical bundle of curves in **X** [20]. By canonical bundle we mean that all curves of the bundle satisfy the Hamilton equations associated with the Hamiltonian function $H$. Having a complete figure provides also an immediate link to SQM through its "statistical interpretation" according to Ballentine [21]. In facts, Eq. (2) can be seen as the continuity equation along the canonical bundle of the positive measure $d\mu = \rho\sqrt{g}d^n q$. Then, if this measure is normalizable, we may retain a statistical picture of the theory as made in the SQM based on the Schrödinger equation or on the Klein-Gordon equation [14–16]. The relationship between the CQG and the SQM is better shown by introducing the scalar wave function

$$\Psi(q,t) = \sqrt{\rho(q,t)} e^{i\frac{S(q,t)}{\hbar}} \tag{6}$$

Inserting the *ansatz* (6) into Eqs. (1) and (2), it can be shown that these nonlinear equations reduce to a linear wave equation for $\Psi(q,t)$. This equation is similar to Schrödinger's equation with additional potential given by the Riemann curvature $R_g$ and is invariant under Weyl's gauge. The full equivalence of the CQG with the SQM was demonstrated in previous works (*9–16*). Because $\rho$ and $S$ are single-valued in the region **X**, so is also for the function $\Psi$.

### 3. The single spin

All the above considerations apply to any quantum system. We specialize now the CQG to particles with spin. Let us start with a single spinning particle. The configuration space is $R^3 \times SO(3)$, the direct product of the Euclidean 3D space and the space of the orientations of the frame attached to the particle. The SO(3) space has three parameters, so that the region **X** we are considering is six dimensional. We may use any set of three parameters in SO(3), as for instance the most convenient Euler's angles $\{\alpha,\beta,\gamma\}$. In these coordinates the action $S$ becomes an angle action, so that on the right of Eq. (4) we may replace zero by any integer multiple of $2\pi$. A closer inspection in Eqs. (1), (2) and (3) shows that the Euler angle $\gamma$ does not appear explicitly in the equations neither in the metric nor in the external fields $a_i$ and $V$. The coordinate $\gamma$ can be then considered as "ignorable" and we may seek for solutions of Eq. (1) having the form

$$S(\alpha,\beta,\gamma,\mathbf{r},t) = \hbar s \gamma + S_0(\alpha,\beta,\mathbf{r},t) \tag{7}$$

with $\mathbf{r} = \{x,y,z\}$ and constant $s$. It can be easily shown that the derivatives $s_z = \partial_z S$ and $s_\zeta = \partial_\gamma S$ are the components of the frame angular momentum along the laboratory $z$ axis and along the particle frame $\zeta$-axis, respectively. We call $s_\zeta$ the *intrinsic helicity* of the particle, as said. Then Eq. (7) shows that the intrinsic helicity takes the constant value $s$. No fields are known to be able to change the value of the intrinsic helicity, so once $s$ is given for a particular kind of particle, it is forever. We may then consider the intrinsic helicity as a constant property of the particle, like charge and mass. Without loss of generality we may always orient the $\zeta$ axis so to have $s > 0$ for all particles. In Eq. (7), $s$ can take any value. However, it can be shown that the requirement of having a normalizable measure $d\mu = \rho\sqrt{g}d^n q$ implies that $s$ can be only integer or half integer multiple of $\hbar$, so that coincides with the quantum spin of the particle. Moreover, the spin component $s_z$ along the laboratory $z$-axis can take only values in the interval $-\hbar s \leq s_z \leq \hbar s$. From the Lagrangian (5) it can be easily shown that the time derivative $d\gamma/dt$ of the ignorable angle $\gamma$ is given by[3]

---

[2] If the region **X** of the configuration space is multiply connected and the loops belongs to different homotopy classes, the loop in Eq. (4) must belong to the trivial topology, i.e, must be contractible to a point. Correspondently, we take only the principal value of $S$, i.e. the value in the sheet of the covering space where the identity is mapped into the identity of **X**.

[3] The function $\gamma(t)$ in Eq. (8) is essentially arbitrary and must not obey Hamilton's equations.

$$\frac{d\gamma}{dt} = \frac{(s_\zeta - s_z \cos\beta)}{m\lambda^2 \sin^2\beta} \tag{8}$$

where $\lambda$ is a constant of the order of the particle Compton wavelength. Because $s_z$ is bounded to $-\hbar s \leq s_z \leq \hbar s$ and $s_\zeta = \hbar s$, we see that $d\gamma/dt \geq 0$ along all admissible paths in the region $\mathbf{X}$ of the configuration space. This remarkable result shows that the spinning particle behaves as a sort of ratchet gyroscope with internal mechanism preventing clockwise rotations around its proper axis. Only counter clockwise rotation is allowed. This is precisely what was assumed by Arthur Jabs in its approach to the SSC [6]. The function $S_0$ in Eq. (7) obeys the HJE obtained from Eq. (1) by replacing $\partial_\gamma S$ with $\hbar s$ so that $S_0$ is the reduced action governing the motion in the subspace $\mathbf{X}_s \subseteq \mathbf{X}$ with $s$ = const.. In our case, the subspace $\mathbf{X}_s$ is the quotient space $\mathbf{X}/SO(2)$ where frames rotated around $\zeta$ of an angle $\gamma$ are considered equivalent. This space is spanned by the two angular coordinates $\{\alpha,\beta\}$ and is isomorphic to the Bloch sphere $S^2$. This is consistent with the physical interpretation of the quantum spin as magnetic moment $\boldsymbol{\mu}$ of the particle, as made in the SQM. In facts, as we shall see, all quantum effects of the SQM are encoded in the reduced action $S_0$; the intrinsic helicity and the corresponding $\gamma$ angle are extraneous to the SQM. If we take the laboratory $z$-axis as quantization axis, the SQM describes the spin $s$ by means of $(2s + 1)$-component spinor $\psi^\sigma(\mathbf{r},t)$ with $\sigma \equiv s_z/\hbar = (-s,-s+1,\ldots,s)$. The connection of the CQG with spinors is made more evident if we consider the scalar wave function $\Psi$ that, in this case, takes the form

$$\Psi(\alpha,\beta,\gamma,\mathbf{r},t) = e^{is\gamma}\Phi(\alpha,\beta,\mathbf{r},t) = e^{is\gamma}\sqrt{\rho(\alpha,\beta,\mathbf{r},t)}e^{i\frac{S_0(\mathbf{r},\alpha,\beta,t)}{\hbar}} \tag{9}$$

A shown in previous works [11,13], the reduced wave function $\Phi$ in this equation is a linear combination of the components of the spinor $\psi^\sigma(\mathbf{r},t)$ whose coefficients depends on the angles $\alpha$ and $\beta$ only. Because the SQM assumes that quantum spin is fully described by the spinor $\psi^\sigma(\mathbf{r},t)$, it is clear from Eq. (9) that the SQM ignores the intrinsic helicity $s$ and the angle $\gamma$ which, on the contrary, fit very naturally in the CQG context. Indeed $s$ and $\gamma$ play no role in the quantum dynamics of spin, in spite of being at the basis of SSC observed in Nature. In this sense, we may consider that, at least in the present context, the CQG is as a realistic *completion* to SQM. Before to considering the case of identical spinning particles, we want to stress that an important difference exists between the wave functions $\Psi$ and $\Phi$. It is well known that in the case of half-integer spin the spinor $\psi^\sigma$ is double-valued, because $\pm\psi^\sigma$ represent identical physical situations after $2\pi$ rotation. On the contrary, the wave function $\Psi$ of the CQG must be single-valued as said above. This last condition, however, poses no constraint on the spinor $\psi^\sigma$, because when $\psi^\sigma$ changes sign also $\Phi$ change sign and this can be always compensated by a (counter clockwise) $2\pi$ rotation of the angle $\gamma$. Because $\gamma$ is ignored in the SQM, its $2\pi$ rotation does not affect the quantum effects of spin, but ensures that $\Psi$ (and $S$) resumes its initial value so that the complete figure of the calculus of variation can be obtained. Then the Lagrangian line integral has a strong minimum. In other words, we use the rotation of the angle $\gamma$ to "unwrap" the closed path in Eq. (4) so to reduce it into the trivial homotopy class (of the two possible homotopies of SO(3)). However, this path-unwrapping procedure can be avoided without loss of generality by considering only motions of the particle frame where complete turns do not occur. More precisely, we will assume henceforth that only motions are considered where the changes $\Delta\gamma$ and $\Delta\alpha$ of the Euler's angles $\gamma$ and $\alpha$ are in the range $0 \leq \Delta\gamma < 2\pi$, and $|\Delta\alpha| < 2\pi - \Delta\gamma$, respectively.

### 4. The Spin-Statistics Theorem

Let us consider now the case of $N$ identical particles with spin. The system configuration space is the direct product $\mathbf{X}\times\mathbf{X}\times\ldots\times\mathbf{X} = \mathbf{X}^N$ where configurations obtained by particle permutation are considered equivalent and mapped in one point. We can denote this space as the quotient space $\mathbf{X}^N/\mathbb{S}_N$, where $\mathbb{S}_N$ is the group of permutations of $N$ objects. A simple inspection in Eqs. (1) and (2) shows that even in this case and in the presence of external fields all angles $\gamma_a$ related to the single particle $a$ ($a = 1,\ldots,N$) are ignorable coordinates. Then, because the intrinsic helicities of identical particles are equal by definition, the action $S$ can be expressed as follows:

$$S(q_1,\ldots,q_N,t) = \hbar s(\gamma_1 + \ldots \gamma_N) + S_0(\tilde{q}_1,\ldots,\tilde{q}_N,t) \tag{10}$$

where $s$ is the common intrinsic helicity of the $N$ particles and $\tilde{q}_a$ are the coordinates of the particle $a$ with $\gamma_a$ removed. Let us consider now a closed path in $\mathbf{X}^N/\mathbb{S}_N$ which ends up with the exchange of the coordinates of two particles, $a$ and $b$, say, while the coordinates of all other particles remain fixed. Because the system recovers its initial configuration this "exchange path" is closed in $\mathbf{X}^N/\mathbb{S}_N$. Then, according to condition (4) for single valued action $S$, we must have

$$\Delta S_0 = \oint_{a \leftrightarrow b} dS_0 = -\hbar s \oint_{a \leftrightarrow b} d(\gamma_1 + ... \gamma_N) = -\hbar s \oint_{a \leftrightarrow b} d(\gamma_a + \gamma_b) = -\hbar s (\Delta \gamma_a + \Delta \gamma_b) \qquad (11)$$

where the notation means that the closed path ends up with the exchange of particles $a$ and $b$. We notice that the reduced action $S_0$ of the system can be eventually multi-valued. In fact, the condition (11) is not trivial because when the particles are identical, the space spanned by the angles $\gamma_a$ and $\gamma_b$ has the topology of the Möbius strip [23] and, furthermore, as said above, clockwise rotations around the proper axes $\zeta_a$ and $\zeta_b$ of the two particles are forbidden. As shown in Fig. 1, the only acceptable closed exchange path is the one circling once around the Möbius strip without intersecting its boundary. Evaluation of the integral (11) along this path yields $\Delta \gamma_a + \Delta \gamma_b = 2\pi$ so that under the exchange of two particles of the system the reduced action must change of $\Delta S_0 = -2\pi\hbar s$. We may extend this result to close paths where more than two particles are exchanged. Because any permutation $p$ can be obtained by a finite number $k_p$ of simple transpositions, we may form such a closed path by concatenating $k_p$ closed path exchanging two particles at a time. Each one of these paths add a term $-2\pi\hbar s$ to $\oint dS_0$ so that, after the permutation $p$, $S_0$ is incremented by $\Delta S_0 = -2\pi\hbar k_p s$. In the SQM the system of $N$ identical particles is described by a higher-order spinor $\psi^{\sigma_1...\sigma_N}(\mathbf{r}_1,...,\mathbf{r}_N,t)$. The connection with the CQG is made through the scalar wave function $\Psi$ that, in this case, has the form (cfr. Eqs. (9) and (10))

$$\Psi(q_1,...,q_N,t) = e^{is\sum_a \gamma_a} \Phi(\tilde{q}_1,...,\tilde{q}_N,t) \qquad (12)$$

Since $S_0 = \arg(\Phi)$, the single valuedness of $\Psi$ required by CQG in a particle permutation $p$, requires that the reduced wave function $\Phi$ changes as $\Phi \to \Phi_{a \leftrightarrow b} = e^{2\pi i k_p s}\Phi = (-1)^{2k_p s}\Phi$. As in the case of the single spin, the function $\Phi$ is a linear combination of the spinor components $\psi^{\sigma_1...\sigma_N}(\mathbf{r}_1,...,\mathbf{r}_N,t)$ with coefficients which depend only on the Euler's angles $\alpha$ and $\beta$ of the particle frames, viz.

$$\Phi(\alpha_1,\beta_1,\mathbf{r}_1,...,\alpha_N,\beta_N\mathbf{r}_N,,t) = \sum_{\sigma_1=-s}^{s} ... \sum_{\sigma_N=-s}^{s} c_{\sigma_1...\sigma_N}(\alpha_1,\beta_1,...,\alpha_N,\beta_N)\psi^{\sigma_1...\sigma_N}(\mathbf{r}_1,...,\mathbf{r}_N,t) \qquad (13)$$

We note that, because the $\sigma$'s are dummy indices, any permutation of the *coordinates* in the function $\Phi$ on the left of Eq. (13) is equivalent to the same permutation on the particle *labels* in the term in the sum on the right, i.e. to the permutation of both coordinates and spin. Now, a direct calculation based in Eqs. (1) and (2) shows that the coefficients $c$'s in the expansion (13) remain unchanged for any permutation of the particle labels. Then, to have the desired behaviour of $\Phi$ under coordinate permutation, we are forced to choose the space-time dependent part of the solution of Eqs. (1) and (2), i.e. the spinor $\psi^{\sigma_1...\sigma_N}(\mathbf{r}_1,...,\mathbf{r}_N,t)$, so that, in a permutation $p$ of the particle labels, $\psi^{\sigma_1...\sigma_N}(\mathbf{r}_1,...,\mathbf{r}_N,t)$ changes by a phase factor $(-1)^{2k_p s} = (\pm 1)^{k_p}$, the double sign $\pm$ depending on being $s$, the common intrinsic helicity of the identical particles, integer or half integer. The only possible choice is to take

$$\psi^{\sigma_1...\sigma_N}(\mathbf{r}_1,...,\mathbf{r}_N,t) = \frac{1}{\sqrt{N!}} \sum_{\alpha=1}^{N!} (-1)^{2sk_\alpha} p_\alpha (\psi_1^{\sigma_1}(\mathbf{r}_1,t)\psi_2^{\sigma_2}(\mathbf{r}_2,t)...\psi_N^{\sigma_N}(\mathbf{r}_N,t)) \qquad (14)$$

where the sum is extended to all permutations of the $N$ particles, in accordance with Pauli's Spin-Statistics Connection.

## 5. Conclusions

In conclusion, we presented a proof of the SSC in the nonrelativistic limit and without making recourse to the quantum field approach. In the CQG framework, this is made possible by the role played by the intrinsic helicity $s_\zeta$ and by the peculiar kinematical constraint on the rotation angle $\gamma$ conjugated to $s_\zeta$. We may regard the intrinsic helicity as a constant property of any elementary particle, which is not considered in the SQM. Therefore, the CQG theory completes SQM, in Einstein's sense. In the CQG, the SSC theorem is a consequence of the allowed values of $s_\zeta$, of the kinematical constraint $d\gamma/dt \geq 0$ and of the requirement to have single valued functions $S$ and $\rho$. Most important, it follows that the CQG handles bosons and fermions on the same footing by a <u>unique scalar wave function $\Psi$</u>. It is the *space* part of this wave function, contained in the spinor fields inside the reduced wave function $\Phi$, which behaves differently under exchange of identical particles. Nowadays the only physical effect due to the intrinsic helicity appears to be the "Pauli's exclusion principle", as shown here, but there is no reason of principle which forbids $s_\zeta$ to be measured, e.g. by a neutron interferometry experiment. However, the realization of such experiment should be difficult because no fields are known which actively interact with the intrinsic helicity of the elementary particles.

**References**

[1] I. Duck, E. C. G. Sudarshan, and Arthur S. Wightman, Am. J. Phys. **67**, 742 (1999).
[2] R. P. Feynman, R. B. Leighton, and M. L. Sands, *The Feynman Lectures on Physics* (Addison-Wesley, Redwood City, Calif., 1989).
[3] L. De Broglie and J. L. Andrade e Silva, *La réinterprétation de la mécanique ondulatoire* (Gauthier-Villars, 1971).
[4] A. P. Balachandran, A. Daughton, Z.-C. Gu, R. D. Sorkin, G. Marmo, and A. M. Srivastava, Int. J. Mod. Phys. A **08**, 2993 (1993).
[5] H. Bacry, Am. J. Phys. **63**, 297 (1995).
[6] A. Jabs, Found. Phys. **40**, 776 (2010).
[7] A. Jabs, Found. Phys. **40**, 793 (2010).
[8] A. S. Davydov, *Quantum Mechanics*, 2 edition (Pergamon Pr, Oxford; New York, 1976). The operator $\hat{s}_\zeta = -i\hbar \partial_\gamma$ is well-known in the group theory, where it is used, together with the spin operator $\hat{s}_z$, to introduce the finite dimensional representations of the rotation group. $\hat{s}_\zeta$ can be identified with the component of the angular momentum of the rotating frame along its proper $\zeta$-axis. Moreover, $\hat{s}_\zeta$ commutes with any one of $\hat{s}_x$, $\hat{s}_y$, $\hat{s}_z$, so that it is an independent observable. However, $\hat{s}_\zeta$ does not enter in the Schrödinger-Pauli and Dirac equations, so that it is actually ignored in SQM and in quantum experiments with spinning particles.
[9] E. Santamato, ArXiv08083237 Quant-Ph (2008).
[10] E. Santamato and F. De Martini, Int. J. Quantum Inf. **10**, 1241013 (2012).
[11] E. Santamato and F. D. De Martini, Found. Phys. **43**, 631 (2013).
[12] F. De Martini and E. Santamato, EPJ Web Conf. **58**, 01012 (2013).
[13] F. De Martini and E. Santamato, Int. J. Theor. Phys. (2013).
[14] E. Santamato, Phys. Rev. D **29**, 216 (1984).
[15] E. Santamato, J. Math. Phys. **25**, 2477 (1984).
[16] E. Santamato, Phys. Rev. D **32**, 2615 (1985).
[17] E. Santamato, Phys. Lett. A **130**, 199 (1988).
[18] H. Weyl, *Space, Time, Matter*, 4th ed. (Dover Publications, Inc., New York, 1952).
[19] F. De Martini and E. Santamato, ArXiv14062970 Quant-Ph (2014).
[20] H. Rund, *The Hamilton-Jacobi Theory in the Calculus of Variations* (Robert E. Krieger, Huntington, N. Y., 1973). Necessary and also sufficient conditions for strong minimum are a) *S* is single valued; b) *S* obeys the HJE; c) the metric $g_{ij}$ is positive definite. In our case all three conditions are met.
[21] L. E. Ballentine, Rev. Mod. Phys. **42**, 358 (1970).
[22] H. Goldstein, *Classical Mechanics* (Addison-Wesley Pub. Co., 1980).
[23] J. M. Leinaas and J. Myrheim, Il Nuovo Cimento B Ser. 11 **37**, 1 (1977).

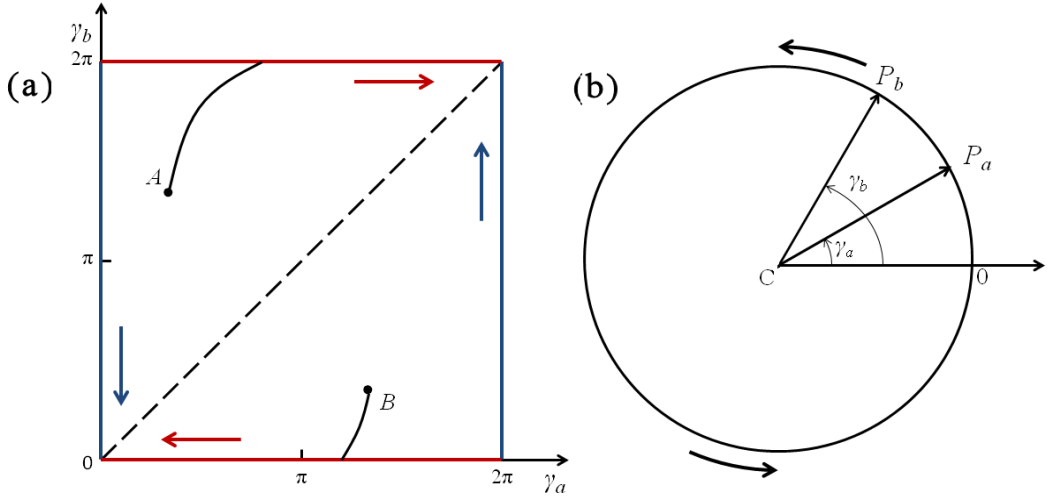

Fig. 1. The exchange path in the plane $(\gamma_a,\gamma_b)$ of the rotation angles of particles $a$ and $b$ along their respective proper axes $\zeta_a$ and $\zeta_b$. We may imagine $\gamma_a$ and $\gamma_b$ as the polar angles of two points $P_a$ and $P_b$ moving on a circle (Fig. 1b). Because 0 and $2\pi$ refer to the same position of $P_a$ and $P_b$, the opposite edges (the blue and the red ones in Fig. 1a) should be identified in a single point in the configuration space. After this identification, the square in Fig. 1a is transformed into a torus. In the case of two identical particles, also the points $(\gamma_a,\gamma_b)$ and $(\gamma_b,\gamma_a)$ should be identified. These points are disposed symmetrically with respect to the diagonal $\gamma_a = \gamma_b$, as points $A$ and $B$ in Fig. 1a. This further identification orients the opposite edges of the square in opposite directions, as shown by the arrows, which transform the torus into a Möbius strip. The dotted diagonal in the figure is the boundary of the strip. Paths joining points like $A$ and $B$ produce the exchange of the two particles. If a path intersects the boundary of the Möbius strip, one of the points $P_a$ and $P_b$ overtakes the other. The kinematical conditions $d\gamma_a/dt \geq 0$ and $d\gamma_b/dt \geq 0$ select only paths with positive slope. This prevents, for example, the paths joining $A$ and $B$ directly. It can easily shown that the only path joining $A$ and $B$ with positive slope and so that no one of the particle frames complete a full $2\pi$ turn ($0 \leq \Delta\gamma_a, \Delta\gamma_b < 2\pi$) is the one shown in the figure (or any other homotopically equivalent to it with positive slope). This path corresponds to one complete turn along the Möbius strip without intersecting the boundary (the point $P_a$ does not overtake $P_b$ in Fig. 1b) and along such path we have $\Delta\gamma_a + \Delta\gamma_b = 2\pi$ as said in the text.